\documentstyle[prl,aps]{revtex}
\begin{document}
\title{Modelling Molecular Motors as Folding-Unfolding Cycles}
\author{Alex Hansen\footnote{Permanent address: 
Department of Physics,
Norwegian University of Science and Technology, N--7034 Trondheim, 
Norway.}, Mogens H. Jensen, Kim Sneppen 
and Giovanni Zocchi}
\address{ Niels Bohr Institute and NORDITA, 
Blegdamsvej 17, DK-2100 {\O}, Denmark}
\date{\today}
%
\maketitle
\begin{abstract}
We propose a model for motor proteins 
based on a hierarchical Hamiltonian that we have previously introduced to
describe protein folding.
The proposed motor model has high efficiency and is consistent
with a linear load-velocity response.
The main improvement with respect to previous models is that this 
description suggests a connection between folding and function
of allosteric proteins.
\end{abstract}

An important class of proteins are alosteric: part of the molecule
undergoes large conformational changes as a result of a chemical reaction which 
takes place in another region of the protein.
One example of this is motor proteins which through a 
series of conformational changes
perform directed motion using the chemical energy gained $ATP$ hydrolysis.
Physicists have proposed that this may be viewed as
a ratchet-like mechanism, similar to Feynman's famous ratchet \cite{Feynman}
but with the addition that ATP binding/hydrolysis
has the ability to turn this ratchet on and off. 
Several models have been proposed along this lines, see
\cite{SPO92,Simon,adjari,prost,julicher,julicher2,DLH,DLH71}.

The energy associated with the chemical cycle of real
motor proteins is comparable to the total binding 
energy of a protein.
This leads us to speculate that the directed
motion associated to motor action may be connected to the
pathway for protein folding.
Thus, we propose to describe the motor variables in terms of 
a pathway parametrization of protein folding \cite{hjsz,hjsz1}
using three variables $\varphi_0$, $\varphi_1$ and $\varphi_2$, 
that each takes value $0$ or $1$.  These variables, which label
the conformations visited during the motor cycle, 
also label subsequent states along the folding
pathway of the protein.
Thus we propose that the energy function of the motor protein
in the absence of ATP and external load is described
by the Hamiltonian \cite{hjsz}:
\begin{equation}
\label{ham0}
H_0 = - \varphi_0 - \varphi_0 \varphi_1 - \varphi_0 \varphi_1 \varphi_2 \;.
\end{equation}
The ground state of $H_0$ is $(\varphi_0,\varphi_1,\varphi_2)=(1,1,1)$.
At low temperatures $T$ the variables will tend to freeze 
into the ground state, whereas they melts for $T>T_c=1/\ln(2)$.
In order to take into account ATP hydrolysis and an external 
load we suggest the following extension of the energy function:
\begin{equation}
\label{hamilton}
H_1 = -[ATP] \; \varphi_0 \; -\; 
\varphi_0 - \varphi_0\varphi_1 - \varphi_0\varphi_1 \varphi_2 
\;+\; F(\varphi_0, \varphi_1, \varphi_2)
\end{equation}
where [ATP] represents the effects of ATP on the chemical potential
of the molecule. [ATP] can cycle between two values, 
say $0$ and $-A$ where $A$ is the energy associated to hydrolysis, 
and $F$ is the external load on the motor. 
With this structure 
we single out one variable $\varphi_0$ to represent
ATP binding and hydrolysis while the two other variables
represent events associated to attachment 
($\varphi_1$) and mechanical motion ($\varphi_2$)
of the motor.

The dynamics of the motor is generated 
by switching ATP between $-A$ and $0$.
When ATP is zero, 
$\varphi_0$ is driven to $\varphi_0=1$ by the first term in the 
energy function. Subsequently this sets an ordering where typically
first the variable $\varphi_1$, then the variable $\varphi_2$
takes the value $1$. 
Thus first the ``leg'' binds to the substrate, then it performs a 
step (or ``power stroke'') $\varphi_2=0 \rightarrow 1$, and it 
advances forward by one step.
In case it  by chance do the opposite, 
i.e. if $\varphi_2$ moves before $\varphi_1$
it represent the event when the stroke takes place before the binding, 
and we assign 
it a one step backwards movement. When $(\varphi_1,\varphi_2)=(1,1)$ 
we initiate the chemical switch [ATP] $ = 0 \rightarrow - A$ which subsequently
forces $\varphi_0 \rightarrow 0$, and accordingly release the binding 
constraints of $\varphi_1,\varphi_2$. 
The ordering of the melting of these variables is random.
The asymmetry in the freezing --- where the process typically 
proceed in the order $\varphi_0 \to 1$, $\varphi_1 \to 1$ and $\varphi_2 \to 1$
--- and the melting --- which can proceed in any order, 
we refer to as the {\it mirror effect\/} \cite{hjsz},
and lies at the heart of the functioning of this motor. 
The total move table is
\begin{eqnarray}
(\varphi_1 , \varphi_2 ) & \;=\;& (0,0) \rightarrow (1,0) \rightarrow (1,1)
\quad\mbox{defines move}\quad x \rightarrow x+dx\\
(\varphi_1 , \varphi_2 ) & \;=\;& (0,0) \rightarrow (0,1) \rightarrow (1,1)
\quad\mbox{defines move}\quad x \rightarrow x-dx\\
(\varphi_1 , \varphi_2 ) & \;=\;& (1,1) \rightarrow (1,0) \rightarrow (0,0)
\quad\mbox{defines move}\quad x \rightarrow x-dx\\
(\varphi_1 , \varphi_2 ) & \;=\;& (1,1) \rightarrow (0,1) \rightarrow (0,0)
\quad\mbox{defines move}\quad x \rightarrow x+dx
\end{eqnarray}
In terms of the normal nomenclature in the literature on motor molecules,
the moves of Eq.(3) represent respectively the binding and 
the working stroke whereas the moves in Eq. (5) undo these.
The moves of Eq. (6) then represent the optimal
sequence of unbinding and recovery.

In order to study the model numerically, we have chosen to use the following
dynamics:  In each update, either zero or one of the $\varphi$-variables changes
value.  The probability that a given change is made (either $(\varphi_0,
\varphi_1,\varphi_2)\to(1-\varphi_0,\varphi_1,\varphi_2)$,
$(\varphi_0,1-\varphi_1,\varphi_2)$, $(\varphi_0,\varphi_1,1-\varphi_2)$, or 
$(\varphi_0,\varphi_1,\varphi_2)$) is proportional to the Boltzmann factor
associated with the configuration the model is entering into,  
\begin{equation}
P (s_i \rightarrow s_f) \propto e^{-H_1(s_f)/T}\;.
\end{equation}
where $s_i$ and $s_f$ are initial and final states respectively.  As the 
variables are discrete, it is not possible to assign a physical time to 
the process.  However, the amount of ATP that is hydrolyzed (i.e., the number
of times $A$ takes the value $-a$) is a natural clock in this process.  

A more realistic dynamics for this model is based on constructing a Langevin
equation from the Hamiltonian (\ref{hamilton}).  Continuous variables may then
be used and a connection with ``real time" may be made.  The Langevin equation
reads
\begin{equation}
\label{langevin}
\frac{d\varphi_i}{dt} \;=\; - \mu_i \frac{dH_1}{d\varphi_i} +\eta_i(t)
\end{equation}
where $\mu_i$ are mobilities, $\eta_i(t)$ thermal noise
and the variables $\varphi_i$ now has to be continuously varying
\cite{footnote}.
To obtain the real cycle times we in addition have to introduce
the residence times in the different states.
We note that if the limiting step of the dynamics is the ATP binding, 
i.e. at very low ATP concentration,
then the efficiency curves given by the phase space method also represent real
time velocity curves up to timescale that is given by ATP hydrolysis rate.
In the opposite limit, the Langevin dynamics captures
the timescale of the motion through the coefficients $\mu_i$.
Experiments show a roughly linear load-velocity characteristics
\cite{Howard-Gittes}. We can obtain a linear dependence with this 
Langevin dynamics for certain realizations of the force but this feature
appears not to be robust with respect to the functional form of the load.

In Figure \ref{fig1} we show the average 
position of the motor as function of the number of times an 
ATP molecule was hydrolyzed, i.e., that the ATP was switched from $0$ to $-A$.
In the figure we show two cases, 
both for $A=4$, one at a low temperature $T=0.25$ 
(the upper curve) and one at a high temperature, $T=5$.
In each case the corresponding mean trajectories are also shown 
(average over 5000 trajectories). First we note the high efficiency, 
for the $T=0.25$ case. About 95\%
of the hydrolysis events lead to forward motion.
Such a high efficiency is seen in real motor proteins 
at least at low ATP concentrations \cite{Block,Hua}.

Coming back to the drag force we note that its form
has to fulfill certain conditions.
Firstly it should act 
through pushing the system into a particular ``corner"
of configuration space.
Secondly, the drag force should not change the ATP parameter $A$ directly.
However, the reaction rate, set by the term $-[\mbox{ATP}]\varphi_0$,
can be influenced through forcing the dynamical variables $\varphi_0$, 
$\varphi_1$ and $\varphi_2$.  Thirdly, the drag force influences the
motor function through changing the rate of entering the state 
$(\varphi_1,\varphi_2)=(1,1)\to (0,0)$ {\it vs.\/} the 
$(\varphi_1,\varphi_2)=(0,0)\to (1,1)$ rate.  
An explicit realization that fulfills these conditions are
\begin{equation}
\label{forcemodel}
F(\varphi_0,\varphi_1,\varphi_2)=+f(\varphi_0+\varphi_0\varphi_1+
\varphi_0\varphi_1\varphi_2)\;.
\end{equation}
This is not the only possible choice, but given the structure of the 
Hamiltonian (\ref{hamilton}), this is a natural one.  Another possible
choice is simply $F=f(\varphi_0+\varphi_1+\varphi_2)$.  Note that with $f$
positive, the drag force acts against the direction of the motion of the motor.

In figure \ref{fig2} we show the efficiency $p$ versus drag $f$
for different temperatures $T$.
We observe that $p$ decreases as the drag force is increased.
For high temperature the efficiency $p$
decreases approximately approximately linearly with $f$.
At lower temperatures one observes that a very high efficiency 
is maintained even for moderate drag.
At larger drag, the motor stops moving forward. The stalling
at $f=1$ occurs because the energy-minimum of $H_0$, Eq.\ (\ref{ham0}),
which is equal to -3, is destroyed by the contribution of the 
drag term, which is $3f$ when $f=1$. For $f>1$, the state 
$(\varphi_0,\varphi_1,\varphi_2)=(0,0,0)$ has lower energy that the state
$(1,1,1)$, and the ATP-cycling looses its effect. 
The net result is that the motor stalls.  
The question now is, what temperature should we choose in the model
to compare with actual motors. The $ATP$ cycle is known
to generate about $25kT_{room}$ and the total binding energy
af a protein domain is of the same order. Further motor proteins
stall at a drag corresponding to a work per step of about $10kT_{room}$.
In the model the total binding energy is $3$ and the ATP cycle 
generates $4$ units of energy. Thus room temperature corresponds
to a temperature in the model of about $0.2$.
In the discussion above, this is what we mean with low temperature. 
Thus our model predicts high efficiency maintained nearly up to stalling drag.

Finally we would like to remark that if we pull on the motor, i.e., 
employ a negative $f$, the success rate increases for some range of pulls,
until finally the efficiency of the motor drops dramatically.
This drop is because for $f$ large and negative both values of $A$ 
lead to the same ground state.
To be quantitative for the case $A=-4$:
For $[\mbox{ATP}]=0$ the ground state is $(1,1,1)$ for all $f<0$.
For $[\mbox{ATP}]=-4$, then for $f>-1/3$ the ground state is $(0,0,0)$
whereas for $f<-1/3$ the ground state is again $(1,1,1)$.
Thus for $f>-1/3$ the motor switch between the
ground states $(0,0,0)$ and $(1,1,1)$ when the ATP variable cycles. 
However when $f\le -1/3$, the $(1,1,1)$ state becomes the ground 
state for both $[\mbox{ATP}]$ values,
and the ATP-cycling no longer forces the motor to switch between states.  The
result is a drop in efficiency.  In Figure \ref{fig2}, the drop occurs at a
smaller $f$ than $-1/3$, but seems to approach this value 
as the temperature is decreased.

We now discuss how the present model is different from existing ones.
It has previously been proposed that motors could work as ratchets
\cite{SPO92,Simon,adjari,prost,julicher,julicher2,DLH,DLH71}.
Our model is indeed a type of ratchet, it cycles between two states,
one of random diffusion and one of directed motion,
much like the sawtooth ratchet originally proposed by Prost {\it et al.\/} 
\cite{prost}.
However a ratchet mechanism similar to the one studied by Prost {\it et al.\/}
would, dependent on the asymmetry of the potential, 
at most make one positive move at half the ATP hydrolysis events.
The other half of the ATP cycles, it will remain in same potential well,
or maybe even stop.  Thus the maximum efficiency would be $1/2$.
In this letter we instead propose, that forward
motion depends on the relative ordering of two events associated
to two different degrees of freedom. 
This results in a motor which is close to 100\% efficient.

In summa the present paper discusses how ordering of events may be utilized 
to construct a simple motor.
The motor contains three variables $\varphi_0$, $\varphi_1$ and $\varphi_2$.  
However, the hierarchical structure of the Hamiltonian (\ref{hamilton})
is easily extended to any number of variables --- and thereby levels.
The hierarchical nature of the Hamiltonian ensures that each level 
controls all subsequent levels, that is we have an explicit realization
of local control of global motion as seen in allosteric proteins.
Furthermore, the model suggests a connection 
between two seemingly unrelated properties: folding and motor action. 

\begin{figure}
\caption{ Position of motor {\it vs.\/} time (ATP clock)
for two different temperatures:
$T=0.25$ (upper two curves) and $T=5$ (lower two curves).  In each case,
both averages over 5000 samples and a single realiztion
are shown.  The value of the
ATP parameter $A$ was equal to -4. There was no external drag force acting on
the motor.
}
\label{fig1}
\end{figure}
\begin{figure}
\caption{Speed of motor (measured in terms of ATP cycling)
{\it vs.\/} drag force  $f$ averaged over 5000 samples for
different temperatures. The ATP parameter $A$ is set to -4. 
}
\label{fig2}
\end{figure}


\vskip-12pt
\begin{thebibliography}{99}
\bibitem{Feynman}
Feynmann R.\ P., Leighton R.\ B.\ and Sands M., 
{\sl The Feynmann lecturers on Physics\/} Vol.\ 1 
(Addison Wesley, Reading, 1963) pp.\ 46.1--46.4.
\bibitem{SPO92}
Simon S., Peskin C. and Oster G.F., Proc.\ Natl.\ Acad.\ Sci.\ {\bf89} 
(1992) 3770.
\bibitem{Simon}
Simon S. M., and Oster G. F.,
Proc.\ Natl.\ Acad.\ Sci.\ {\bf 89} (1998) 3770.
\bibitem{adjari}
Ajdari A., Mukamel D., Peliti L.\ and Prost, J.,
J.\ Physique I, {\bf 4} (1994) 1551.
\bibitem{prost}
Prost J., Chauwin J.-F., Peliti L.\ and Ajdari A., Phys.\ Rev.\
Lett.\ {\bf 72} (1994) 2652.
\bibitem{julicher}
Julicher F.\ and Prost, J.\  Phys.\ Rev.\ Lett.\ {\bf 75} (1995)
2618.
\bibitem{julicher2}
Julicher F., Adjari A.\ and Prost, J.,  Rev.\ Mod.\ Phys.\ {\bf 69} 
(1997) 1269.
\bibitem{DLH}
Leibler S.\ and Huse D., J.\ Cell.\ Biol.\ {\bf 121} (1993) 1357.
\bibitem{DLH71}
Duke T.\ and Leibler S., Biophys.\ J.\ {\bf 71} (1996) 1235.
\bibitem{hjsz}
Hansen A., Jensen M. H., Sneppen K.\ and
Zocchi G., Physica A {\bf 250} (1998) 355.
\bibitem{hjsz1}
Hansen A., Jensen M. H., Sneppen K.\ and
Zocchi G., Europhys.\ Journ.\ B {\bf 6} (1998) 157.
\bibitem{footnote}
The actual implementation of the Langevin dynamics is slightly awkward.
One have to transform $\varphi_i \rightarrow V(\phi_i)$
with $V(x)=x$ for $x \in [0,1]$ and $V(x)=1$ for $x\in[1,2]$.
This is to avoid that the unfolding transition depends on time step $dt$.
Further one have to introduce lower and upper thresholds for
identifying the states, thus the number of parameters grows considerably.
\bibitem{Howard-Gittes}
Howard J.\ and Gittes F.\ in {\sl Physics of Biological Systems\/}
edited by Flyvbjerg H., Hertz J., Jensen M.\ H., Mouritsen O.\ G.\
and Sneppen K.\ (Springer Verlag, Berlin, 1997).
\bibitem{Block}
Schnitzer, M.\ J.\ and Block, S.\ M., 
Nature {\bf 388} (1997) 386.
\bibitem{Hua}
Hua, W., Young, E.C., Fleming M. L.\ and Gelles, J.,
Nature {\bf 388} (1997) 390.
\end{thebibliography}
\end{document}